\pdfoutput=1
\documentclass[usenatbib]{mn2e}
\usepackage{apjfonts}
\usepackage{amssymb}
\usepackage{amsmath}
\usepackage{ctable}
\usepackage{fixltx2e} 

\newcommand{\be}{\begin{equation}}
\newcommand{\ee}{\end{equation}}

\newcommand{\msun}{M_{\sun}}

\newcommand{\paperone}{Paper {\small I}}
\newcommand{\papertwo}{Paper {\small II}}

\newcommand{\flast}{f_{\ell}}
\newcommand{\Sprime}{S^{\prime}}

\newcommand\plotonesize[2]
 {\centering \leavevmode \includegraphics[width={#2\columnwidth}]{#1}}

\newcommand{\acknowledgments}{\begin{small}\section*{Acknowledgments}\end{small}}
\newcommand\altaffilmark[1]{$^{#1}$}
\newcommand\altaffiltext[1]{$^{#1}$}
\title[CMF Variation in Galaxies]{Variations in the Stellar CMF \&\ IMF: from Bottom to Top
\vspace{-0.5cm}}

\author[Hopkins]{
\parbox[t]{\textwidth}{ 
Philip F. Hopkins\altaffilmark{1}\thanks{E-mail:phopkins@astro.berkeley.edu}} 
\vspace*{6pt} \\
\altaffiltext{1}{Department of Astronomy, University of California
  Berkeley, Berkeley, CA 94720\vspace{-1.1cm}} \\
}

\date{Submitted to MNRAS, March, 2012\vspace{-0.6cm}}
\begin{document}
\maketitle
\label{firstpage}

\begin{abstract}

We use a recently-developed analytic model for the interstellar medium (ISM) structure from scales of giant molecular clouds (GMCs) through star-forming cores to explore how the pre-stellar core mass function (CMF) and, by extrapolation, stellar initial mass function (IMF) should depend on both local and galactic properties. If the ISM is supersonically turbulent, the statistical properties of the density field follow from the turbulent velocity spectrum, and the excursion set formalism can be applied to analytically calculate the mass function of collapsing cores on the smallest scales on which they are self-gravitating (non-fragmenting). Two parameters  determine the model: the disk-scale Mach number $\mathcal{M}_{h}$ (which sets the shape of the CMF), and the absolute velocity/surface density (to assign an absolute scale). We show that, for normal variation in disk properties and gas temperatures in cores in the Milky Way (MW) and local galaxies, there is almost no variation in the high-mass behavior of the CMF/IMF. The slope is always close to Salpeter down to $\lesssim 1\,\msun$. We predict modest variation in the sub-solar regime, mostly from variation in $\mathcal{M}_{h}$, but this is consistent with the $\sim1\sigma$ observed scatter in sub-solar IMFs in local regions. For fixed global (galaxy) properties, there is little variation in shape or ``upper mass limit'' with parent GMC mass or density. However, in extreme starbursts -- ULIRGs and merging galaxy nuclei -- we predict a  {\em bottom-heavy} CMF. This agrees well with the IMF recently inferred for the centers of Virgo ellipticals, believed to have formed in such a process. The CMF is bottom heavy despite the gas temperature being an order of magnitude larger, because $\mathcal{M}_{h}$ is also much larger. Larger $\mathcal{M}_{h}$ values make the ``parent'' cloud mass (the turbulent Jeans mass) larger, but promote fragmentation to smaller scales (set by the {sonic} mass, not the Jeans mass); this shifts the turnover mass and also steepens the slope of the low-mass CMF. The model may also predict a top-heavy CMF for the in-situ star formation in the sub-pc disk around Sgr A$^{\ast}$, but the relevant input parameters are uncertain.

\end{abstract}

\begin{keywords}
star formation: general --- galaxies: formation --- galaxies: evolution --- galaxies: active --- 
cosmology: theory
\vspace{-1.0cm}
\end{keywords}

\vspace{-1.1cm}
\section{Introduction}
\label{sec:intro}

The origin of the stellar initial mass function (IMF) is a question of fundamental importance for the study of star formation, stellar evolution and feedback, and galaxy formation. It is an input into a huge range of models of all of these phenomena, and a necessary assumption when deriving physical parameters from many observations. However, despite decades of theoretical study, it remains poorly understood. 

A particularly important question which has received great attention is whether the IMF is indeed ``universal'' or in fact varies in some environments \citep[for a recent review, see][]{kroupa:2011.imf.review}. A growing body of observations find that the high-mass behavior of the IMF in the MW and nearby Small/Large Magellanic Cloud (SMC/LMC) systems is remarkably uniform, with a Salpeter-like ($\alpha=2.3$ in ${\rm d}N/{\rm d}M\propto M^{-\alpha}$) slope consistently appearing in essentially all directly measured systems: young clusters, globulars, disk, bulge, low/high mass GMCs, etc.\ \citep[see e.g.][and references therein]{kroupa:imf,chabrier:imf,bastian:2010.imf.universality}. There are suggestions of greater variation -- albeit still modest -- in how rapidly the IMF turns over in the sub-solar regime ($\lesssim 1\,\msun$), but the observations remain uncertain at these masses \citep[references above and e.g.][]{chabrier:2005.review2,allen:2005.lowmass.imf.stats,thies:2007.imf.lowmass.discontinuity,deacon:2008.lowmass.disk.imf,covey:2008.lowmass.disk.imf,de-marchi:2010.pdmf.cluster.var}. 

There may be, however, variation in more extreme systems. The IMF in most of the galactic center is ``normal'' \citep{lockmann:2010.gal.center.imf}, but there is compelling evidence that the pair of eccentric young circum-BH nuclear disks at just $\sim 0.1-0.5\,$pc from Sgr A$^{\ast}$ may have a top-heavy IMF \citep[either with a flat slope, $\alpha=1.7-2$, or a sharp sub-solar cutoff; see][]{paumard:2006.mw.nucdisk.imf,maness:2007.mw.nucdisk.imf,
bartko:2010.mw.nucdisk.imf}. These are believed to have formed in-situ, from a circum-BH thin disk (a quasi-accretion disk), as opposed to e.g.\ galactic center star clusters (that formed in their own clouds), and so may have had very different formation conditions from ``normal'' stars. 
There have been a number of suggestions of top-heavy IMFs in high-star formation rate galaxies \citep[e.g.][]{hoversten:2008.imf.var.sdss,gunawardhana:2011.imf.var.w.sfr,marks:2012.lowmetal.topheavy.imf,dabringhausen:2012.ucd.imf} and/or high-redshift galaxies \citep[from studies of galactic evolution; see][]{hopkinsbeacom:sfh,vandokkum:imf.evol,dave:imf.evol}, but these are indirect, model-dependent constraints and subject to alternative interpretations. 
Recently, though, \citet{van-dokkum:2010.bottom.imf.ells,van-dokkum:2011.imf.test.gcs} and \citet{spiniello:2012.bottom.heavy.imf.massive.gal} have attempted to directly constrain the IMF in the centers of Virgo and other nearby elliptical galaxies, and find the observations strongly favor a {\em bottom-heavy} IMF (as steep as $\alpha\sim3$). This is independently suggested by kinematic and lensing data \citep[see e.g.][]{treu:2010.early.type.imf.lensing,auger:2010.early.type.imf.lensing,
cappellari:2012.imf.variation,dutton:2012.bulge.bottomheavy.imf} 
as well as spectral modeling \citep[references above and][]{ferreras:2012.imf.vs.vel.dispersion.early.type.gals}. Since these galaxies formed their stars at high redshifts, this is a powerful constraint on any redshift evolution in the IMF. Moreover, it is well-established that stars in the central regions of ellipticals are formed via inflows of gas to large densities in extreme (probably merger-induced) nuclear starbursts; this is the process seen ``in action'' in the centers of local ULIRGs \citep[see][and references therein]{kormendysanders92,mihos:cusps,hibbard.yun:excess.light,rj:profiles,
hopkins:cusps.mergers,hopkins:cusps.ell}. So this should be directly related to the IMF in starburst environments, which many theoretical models have predicted should be top-heavy -- the opposite of what is observed -- owing to the higher observed gas temperatures and potentially large gas accretion rates \citep[see e.g.][]{padoan:1997.density.pdf,larson:2005.imf.scale.thermalphysics}. 

Explaining the weak IMF variation in ``normal'' systems has actually been a challenge for theoretical models, which often predict large IMF variations with e.g.\ the local thermal Jeans mass \citep[although see][]{jappsen:2005.imf.scale.thermalphysics}. It is a particularly powerful constraint for any model which attempts to predict IMF variations in systems where it has been suggested. There is no simple analytic model which has been shown to naturally explain both the weak IMF variation in MW and local group systems while also predicting large IMF variations in more extreme environments similar to those above. 

In this paper, we therefore attempt to develop an analytic understanding of IMF variation and apply it to these observations. It is increasingly clear that the ISM is governed by supersonic turbulence over a wide range of scales. Numerical and analytic arguments have shown that the density probability distribution function (PDF) in this regime approaches a lognormal distribution \citep{vazquez-semadeni:1994.turb.density.pdf,
padoan:1997.density.pdf,ostriker:1999.density.pdf}. \citet{hennebelle:2008.imf.presschechter} used this to propose a simple theoretical model for the stellar IMF, similar to the derivation of the halo mass function in \citet{pressschechter}: if the PDF is lognormal, one can calculate the fraction of gas above a critical density (essentially the Jeans criterion for thermal+turbulent support) on a given scale, and associate this with the mass function. However, like the \citet{pressschechter} argument, this is only approximate as it does not resolve the ``cloud in cloud'' problem. Moreover, the focus therein was exclusively on small-scale properties, so a number of properties must be simply assumed rather than calculated from first principles and the shape and normalization of the IMF cannot be predicted {\em a priori} from global galaxy properties. Recently, \citet{hopkins:excursion.ism} (hereafter \paperone) and \citet{hopkins:excursion.imf} (\papertwo) generalized this by showing how the mathematical excursion set formalism can be applied to turbulent density fields in galaxy disks, over the entire dynamic range from galactic scales to below the sonic length. This resolves these ambiguities, and in principle allows for a rigorous calculation of the core mass function (CMF), defined as the {\em last-crossing distribution} -- specifically, the mass function of bound objects defined on the smallest scales on which they remain self-gravitating but do not have self-gravitating sub-regions (i.e.\ are not fragmenting). \papertwo\ showed that the resulting CMF agrees well with CMFs observed, and (provided some plausible mean core-to-stellar mass conversion efficiency as in \citealt{matzner:2000.lowmass.sf.eff}, which we stress is not explicitly part of the model) can explain all of the major features of the canonical MW stellar IMF. Moreover, because the model self-consistently includes the scales of the parent disk (where most of the turbulent power is concentrated), the distribution of parent cloud properties left as free parameters in \citet{hennebelle:2008.imf.presschechter} can be derived; \paperone\ showed that the model is completely specified by the assumed turbulent velocity power spectrum and global disk properties (surface density, velocity dispersion). 

In this paper, we use the model for the core MF from \papertwo\ to study how the predicted IMF might vary as a function of galactic properties and across different MW regions. 
In \S~\ref{sec:imf} we summarize the model. In \S~\ref{sec:scalings} we derive some qualitative scalings for how the resulting CMF should vary with global parameters. 
In \S~\ref{sec:extreme} we compare the exact model results for different observed properties of the MW, starburst/ULIRG systems, and the circum-BH MW stellar disk. 
In \S~\ref{sec:var.local} we compare the model predictions for more typical variations in properties across the MW disk and in nearby galaxies, and examine how the CMF should vary within the MW in molecular clouds (MCs) of different masses and regions of high/low densities. 
In \S~\ref{sec:discussion} we summarize our results and discuss the implications of the model.

Before proceeding, we must caution that what the model here predicts directly is the pre-stellar {\em core} mass function, {\em not} the stellar IMF. It is by no means clear that the two can be trivially identified with one another (modulo some mean ``efficiency'' of core-to-stellar mass conversion). In addition to the turbulent gas structure, the processes of fragmentation as cores contract and form protostars, accretion onto those protostars, outflows expelling material from cores, and ejection of protostars before accretion is exhausted, can all change the relation between CMF and IMF. It is beyond the scope of this paper to include all of these physics and therefore to make a rigorous absolute prediction of the IMF. Our intention is instead to study how the CMF varies as a function of global galaxy and turbulent properties. Our qualitative expectation is that otherwise identical CMFs, with similar microphysics, should produce a similar stellar IMF, and more bottom/top-heavy CMFs should correspond to more a bottom/top-heavy IMF. These assumptions are valid if the local star formation process depends only on e.g.\ the global core mass and/or stellar microphysics. But lacking a complete understanding of star formation, it remains possible that these microphysics will serve to ``regulate'' against variations in the IMF in a manner that offsets some of the predictions here.

\vspace{-0.5cm}
\section{The CMF \&\ IMF}
\label{sec:imf}

If density fluctuations in supersonic turbulence are lognormal, 
then the variable $\delta({\bf x})\equiv \ln{[\rho({\bf x})/\rho_{0}]}+S/2$, 
where $\rho({\bf x})$ is the density at a point ${\bf x}$, 
$\rho_{0}$ is the global mean density and $S$ is the variance in $\ln{\rho}$, 
is normally distributed according to the PDF:\footnote{The $+S/2$ term 
in $\delta$ is required so that the integral of $\rho\,P_{0}(\rho)$ correctly 
gives $\rho_{0}$ with $\langle \delta\rangle = 0$.}
\be
P_{0}(\delta\,|\,S) = \frac{1}{\sqrt{2\pi\,S}}\,\exp{\left(-\frac{\delta^{2}}{2\,S} \right)}
\ee
More generally, we can evaluate the field $\delta({\bf x}\,|\,R)$, 
which is the $\delta({\bf x})$ field averaged around the point ${\bf x}$ with 
some window function of characteristic radius $R$; this is also normally distributed, with a 
variance $S(R)$ that is directly related to the density power spectrum. 

\papertwo\ showed that the CMF is equivalent to the last-crossing distribution, given by the numerical solution to the Volterra integral equation:
\begin{align}
\label{eqn:flast}
\flast(S) = g_{1}(S) + \int_{S}^{S_{i}}\,{\rm d}S^{\prime}\,\flast(S^{\prime})\,g_{2}(S,\,S^{\prime})
\end{align}
where 
\begin{align}
g_{1}(S) &= {\Bigl [}2\,\frac{dB}{dS} -\frac{B(S)}{S}{\Bigr]}\,P_{0}(B(S)\,|\,S)\\
g_{2}(S,\,\Sprime) &= {\Bigl[}\frac{B(S)-B(\Sprime)}{S-\Sprime} 
+\frac{B(S)}{S}-2\,\frac{dB}{dS}{\Bigr]}\times\\
\nonumber& P_{0}[B(S)-B(\Sprime)\,(S/\Sprime)\,|\,(\Sprime-S)\,(S/\Sprime)]
\end{align}
and $B(S)$ is the minimum value of the overdensity $\delta({\bf x}\,|\,R)$ which defines objects of 
interest (here, self-gravitating regions).

In \paperone\ we derive $S(R)$ and $B(S)$ from simple theoretical considerations for all scales in a galactic disk. For a given turbulent power spectrum, $S(R)$ is determined  
by summing the contribution from the velocity variance on all scales $R^{\prime}>R$
\begin{align}
\label{eqn:S.R}
S(R) &= \int_{0}^{\infty} 
|W(k,\,R)|^{2}
\ln{{\Bigl [}1 + \frac{3}{4}\,
\frac{v_{t}^{2}(k)}{c_{s}^{2} + \kappa^{2}\,k^{-2}}
{\Bigr]}} 
{\rm d}\ln{k} 
\end{align}
where $W$ is the window function for the smoothing.\footnote{For convenience 
we take this to be a $k$-space tophat inside $k<1/R$, which is implicit in our 
previous derivation, but we show in \paperone\ and \papertwo\ that this has little 
effect on our results.} $B(R)$ is 
\be
B(R) = \ln{\left(\frac{\rho_{\rm crit}}{\rho_{0}} \right)} + \frac{S(R)}{2}
\ee
\begin{align}
\label{eqn:rhocrit}
\frac{\rho_{\rm crit}}{\rho_{0}} \equiv \frac{Q}{2\,\tilde{\kappa}}\,\left(1+\frac{h}{R} \right)
{\Bigl[} \frac{\sigma_{g}^{2}(R)}{\sigma_{g}^{2}(h)}\,\frac{h}{R}  + 
\tilde{\kappa}^{2}\,\frac{R}{h}{\Bigr]} 
\end{align}
where $\rho_{0}$ is the mean midplane density of the disk, $Q=\sigma_{g}[h]\,\kappa/(\pi G\,\Sigma_{\rm gas})$ is the standard Toomre $Q$ parameter,\footnote{In what follows, we will assume $Q\sim1$ in calculating some values; observed and simulated systems seem to uniformly converge to this value independent of the detailed physics (see e.g.\ \citealt{gammie:2001.cooling.in.keplerian.disks,hopkins:fb.ism.prop}; but also \citealt{marks:2010.globular.imf.form}). We stress, though, that the way it enters the equations here is only as a parameter encapsulating the ratio of the ``normal'' density to that required for self-gravity (i.e.\ a virial-like parameter), {\em not} as any statement about whether the global disk is ``in equilibrium'' (the equations above apply as well to a non-steady state system). As a result, unless $Q$ is very large, the effect on the quantities shown here is primarily only on their normalization (suppressing it with larger $Q\propto \rho_{\rm crit}$), which we treat as arbitrary.} 
$\tilde{\kappa}=\kappa/\Omega=\sqrt{2}$ for a constant-$V_{c}$ disk, 
and 
\be
\sigma_{g}^{2}(R) = c_{s}^{2} + \langle v_{t}^{2}(R) \rangle 
\ee 
Here and above $\Sigma_{\rm gas}$ is the disk gas surface density and $h$ the (exponential) disk scale height (which can freely vary with radius). 
The mapping between radius and mass is 
\be
M(R) \equiv 4\,\pi\,\rho_{\rm crit}\,h^{3}\,
{\Bigl[}\frac{R^{2}}{2\,h^{2}} + {\Bigl(}1+\frac{R}{h}{\Bigr)}\,\exp{{\Bigl(}-\frac{R}{h}{\Bigr)}}-1 {\Bigr]}
\ee
It is easy to see that on small scales, these scalings reduce to the Jeans criterion 
for a combination of thermal ($c_{s}$) and turbulent ($v_{t}$) support, with $M=(4\pi/3)\,\rho_{\rm crit}\,R^{3}$; on large scales it becomes the Toomre criterion 
with $M=\pi\Sigma_{\rm crit}\,R^{2}$. The mass function is:
\be
\frac{{\rm d}n}{{\rm d}M} = 
\frac{\rho_{c}(M)}{M}\,\flast(M)\,{\Bigl |}\frac{{\rm d}S}{{\rm d}M} {\Bigr |}
\ee

There are only two parameters that completely specify the model in dimensionless 
units. These are the spectral index $p$ of the turbulent velocity 
spectrum, $E(k)\propto k^{-p}$ (usually $p\approx5/3-2$), 
and its normalization, which we define by the Mach number on large scales 
$\mathcal{M}_{h}^{2}\equiv \langle v_{t}^{2}(h)\rangle/c_{s}^{2}$. 
The dimensional parameters $h$ (or $c_{s}$) and $\rho_{0}$ simply rescale 
the predictions to absolute units.

In \citet{hopkins:excursion.clustering}, we extended the derivation to solve the ``two-barrier'' problem. 
This is the solution for $\flast(S\,|\,\delta_{0}[S_{0}])$, i.e.\ given some value $\delta_{0}$ 
on a larger-scale $S_{0}<S$. We refer there for details, but note that this ultimately reduces to the 
solution for $\flast(S)$ above with the replacements $B(S)\rightarrow B(S)-\delta_{0}$, 
$B(S^{\prime})\rightarrow B(S^{\prime})-\delta_{0}$, 
$S\rightarrow S-S_{0}$, and $S^{\prime}\rightarrow S^{\prime}-S_{0}$.

\vspace{-0.5cm}
\section{Qualitative Scalings}
\label{sec:scalings}

For a choice of $p$ and $\mathcal{M}_{h}$, it is straightforward to numerically 
determine the last-crossing mass function (CMF). 
Unfortunately a closed-form solution is not generally possible. 
However, we show in \papertwo\ that on sufficiently small scales (near/below the 
sonic length), 
the ``run'' in $S(R)\approx S_{0}(R_{\rm sonic})$ becomes small (since most of the power 
contributing in Eq.~\ref{eqn:S.R} comes from large scales), 
while $B(S)$ rises rapidly, so 
${\rm d}B/{\rm d}S \gg B(S)/S \gg 1$. In this limit, the MF can be approximated as 
\be
\label{eqn:approx}
\frac{{\rm d}n}{{\rm d}M}
\sim \frac{\rho_{\rm c}(M)}{M^{2}\sqrt{2\pi S_{0}}}
{{\Bigl |}\frac{{\rm d}\ln{\rho_{c}}}{{\rm d}\ln{M}}{\Bigr |}}\,
\exp{\left[-\frac{(\ln{[\rho_{c}/\rho_{0}]}+S_{0}/2)^{2}}{2\,S_{0}} \right]}
\ee

Since we are interested in the behavior on small scales, we drop higher-order 
terms in $h$ and can re-write $\rho_{\rm crit}$ in terms of the sonic length 
\be
R_{\rm sonic}\equiv R[v_{t}^{2}=c_{s}^{2}] = h\,\mathcal{M}_{h}^{-2/(p-1)}
\ee
which for $p=2$ becomes just 
\be
\frac{\rho_{\rm crit}}{\rho_{0}} = \frac{\mathcal{M}_{h}^{2}}{\sqrt{2}}\,
\left(\frac{R_{\rm sonic}}{R} \right)\,
\left[
1 + 
\frac{R_{\rm sonic}}{R} 
\right]
\ee
with $M \approx (4\pi/3)\,\rho_{\rm crit}(R)\,R^{3}$.
There is clearly a change in behavior below $R\approx R_{\rm sonic}$. 
In \papertwo\ we show that $R \gtrsim R_{\rm sonic}$ 
(where $\rho_{\rm crit} \propto M^{-1/2}$)
corresponds to the 
high-mass end of the CMF/IMF, giving a nearly power-law behavior, 
while $R \ll R_{\rm sonic}$ ($\rho_{\rm crit} \propto M^{-2}$)
corresponds to the low-mass end, where there is a 
quasi-lognormal turnover in the CMF and IMF. 

The turnover mass is related to $M_{\rm sonic}\equiv M(R_{\rm sonic})$, 
\be
\label{eqn:msonic}
M_{\rm sonic} = \frac{2}{3}\,\frac{c_{s}^{2}\,R_{\rm sonic}}{G}
= \frac{2\sqrt{2}\,Q^{-1}}{3\pi}\,\frac{c_{s}^{4}}{G^{2}\,\Sigma_{\rm gas}}
\ee
where in the second equality we use the standard Toomre $Q$ parameter to relate this to global 
parameters.

Note that this is the global thermal Jeans mass in a disk -- not, actually, 
the local Jeans mass $\propto c_{s}^{3}\,G^{-3/2}\,\rho^{-1/2}$, 
which is dimensionally the same but differs by powers of $\mathcal{M}_{h}$ and $\rho/\rho_{0}$. 
Both, however are different from the turbulent Jeans mass, 
\be
M_{\rm turb,\, Jeans} \approx \frac{\sigma_{g}^{4}(h)}{\pi\,G^{2}\,\Sigma_{\rm gas}}
\ee
We showed in \paperone\ that $M_{\rm turb,\, Jeans}$ corresponds to the characteristic mass of 
MCs/GMCs.

From Eq.~\ref{eqn:approx}, when $R\gg R_{\rm sonic}$, the the high-mass slope of the CMF 
(${\rm d}n/{\rm d}M \propto M^{-\alpha}$) is approximately
\be
\alpha_{\rm high} \approx \frac{3\,(1+p^{-1})}{2} 
+ \frac{(3-p)^{2}\,\ln{(M/M_{0})}-p\,\ln{2}}{2\,S(M)\,p^{2}} 
\ee
where $M_{0}=\rho_{0}\,h^{3}$ and 
the second term (in $S_{0}$) is small for the reasonable mass range ($\lesssim 0.1$), 
so we find 
\be
\label{eqn:slope.highmass}
\alpha_{\rm high}  \approx \frac{3}{2}\,(1+p^{-1})
\ee
nearly {\em independent} of Mach number $\mathcal{M}_{h}$. The exact 
solution reproduces this but with an even weaker $p$ dependence (\papertwo).

On the other hand, the low-mass slope is 
\begin{align}
\alpha_{\rm lowmass}\approx 3 -
\frac{1}{S}\left(\ln{2}-4\ln{\mathcal{M}_{h}-2\ln{(M/M_{\rm sonic})}} \right)
\end{align}
which for typical parameters (using 
the approximate scaling $S_{0}\propto\ln{\mathcal{M}_{h}^{2}}$) 
reduces to 
\be
\label{eqn:slope.lowmass}
\alpha_{\rm low}\approx 1.9 + \frac{0.6}{\ln{\mathcal{M}_{h}}}\,\ln{(M/M_{\rm sonic})}
\ee

\begin{figure}
    \centering
    \plotonesize{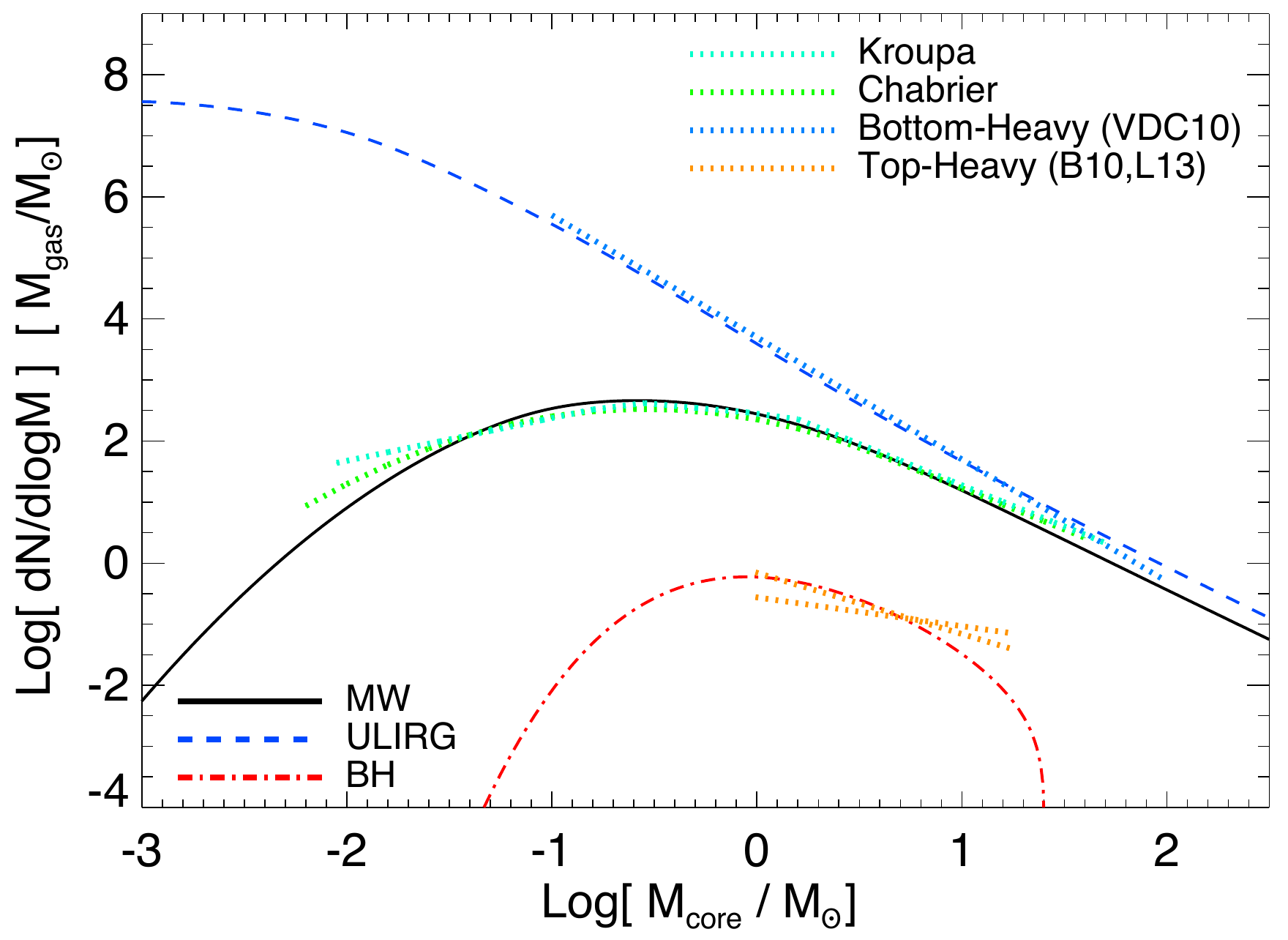}{0.95}
    \caption{The exact predicted last-crossing mass function -- i.e.\ star-forming core MF -- 
    from Eq.~\ref{eqn:flast} for the properties of different observed systems. 
    The predicted CMF mass scale is set by the sonic mass (Eq.~\ref{eqn:msonic}); 
    because of the assumption of marginal disk stability and a turbulent cascade, 
    the shape is entirely specified by the dimensionless Mach number on disk scales $\mathcal{M}_{h}$. 
    {\em MW:} Prediction for canonical local parameters: core minimum temperature 
    $T=10\,$K, disk gas surface density $\Sigma_{\rm gas}\sim10\,\msun\,{\rm pc^{-2}}$, 
    and global turbulent velocity dispersion $\sigma_{\rm g}(h)\sim 10\,{\rm km\,s^{-1}}$. 
    We compare the observed \citet{kroupa:imf} and \citet{chabrier:imf} IMFs, 
    shifted in mass by an assumed star-to-core formation efficiency $M_{\ast}/M_{\rm core}=0.5$ \citep{matzner:2000.lowmass.sf.eff}. 
    {\em ULIRG:} Prediction for properties typical in the central $\sim$kpc of ULIRGs and mergers, 
    the stellar remnants of which dominate the central light in most ellipticals: 
    $T\sim 65\,$K, $\Sigma_{\rm gas}\sim3\times10^{3}\,\msun\,{\rm pc^{-2}}$, 
    $\sigma_{\rm g}(h)\sim 80\,{\rm km\,s^{-1}}$. We compare the bottom-heavy IMF 
    (slope $=-3$ from $0.1-100\,\msun$) fit to the centers of ellipticals in 
    \citet{van-dokkum:2010.bottom.imf.ells}. 
    {\em BH:} Prediction for properties of the MW $\sim$pc-scale circum-BH disk 
    (these are less well-determined, see text). We compare a top-heavy IMF 
    (slope $=-1.7\pm0.2$, the range in e.g.\ \citealt{bartko:2010.mw.nucdisk.imf} \&\ \citealt{lu:2013.mw.center.imf}) 
    over the predicted range.
    \label{fig:imf.env}}
\end{figure}

\begin{figure}
    \centering
    \plotonesize{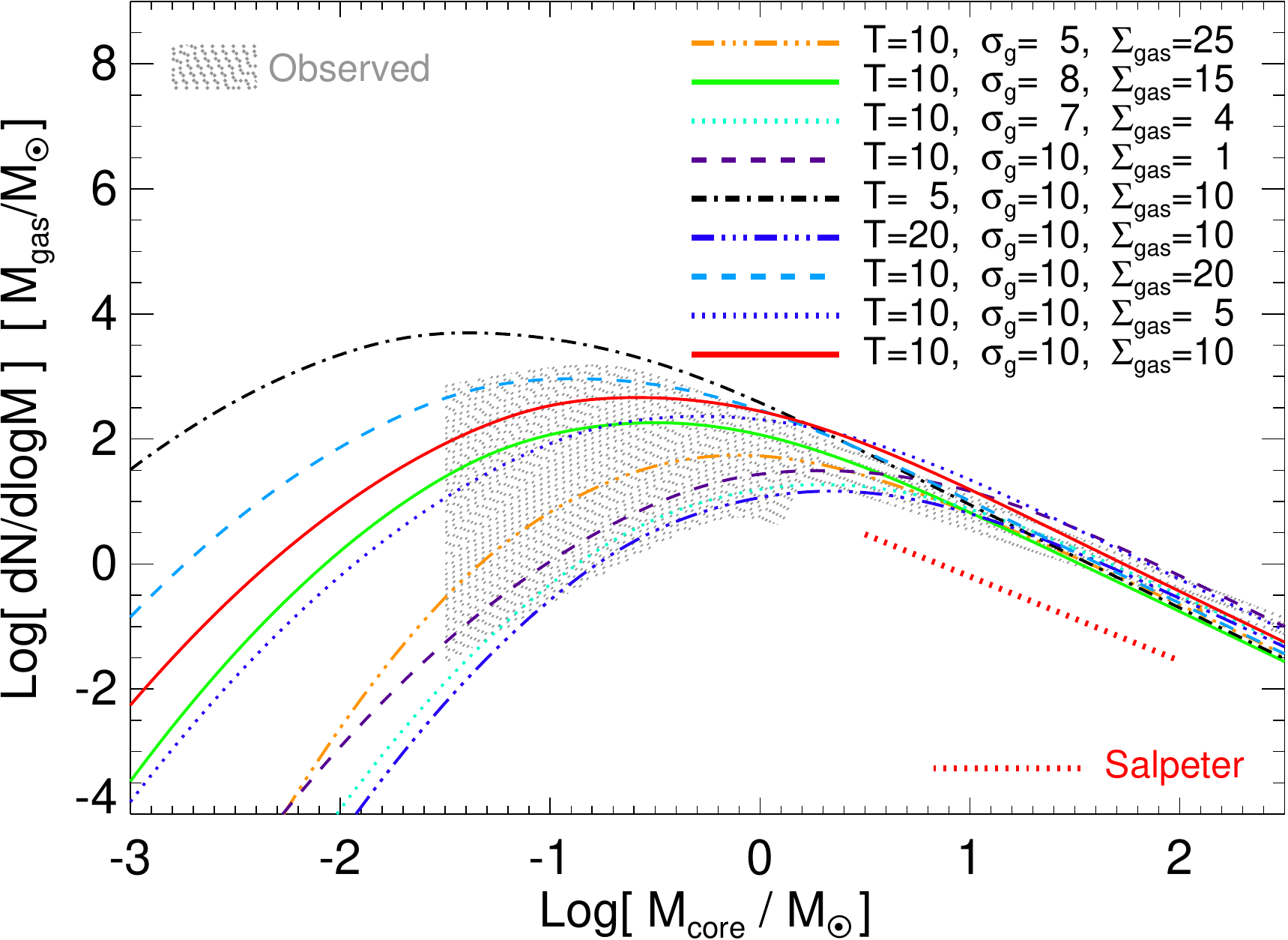}{0.95}
    \caption{The CMF as Fig.~\ref{fig:imf.env}, but for a more ``normal'' range of 
    global parameters typical of the MW and local galaxies. We consider a few cases 
    that vary $\sigma_{\rm gas}$ and $\Sigma_{\rm gas}$ together as expected 
    for an exponential disk with $V_{c}=$constant. We then consider 
    independent variations of $T$, $\sigma$, $\Sigma_{\rm gas}$. 
    In all cases the high-mass behavior is nearly identical and close to Salpeter. 
    The low-mass turnover still varies within this range, primarily with $\mathcal{M}_{h}$, 
    but less severely than in Fig.~\ref{fig:imf.env}. We compare (grey shaded) the $\pm1\,\sigma$ range 
    of observed behavior from the IMF and PDMF compilations in     
    \citet{bastian:2010.imf.universality,de-marchi:2010.pdmf.cluster.var}.$^{\ref{foot:bastian.imf.corr}}$
    \label{fig:imf.var.normal}}
\end{figure}

\vspace{-0.5cm}
\section{CMF \&\ IMF Variation in Extreme Systems}
\label{sec:extreme}

We now consider an observationally motivated example.

First, we consider canonical MW parameters: 
a minimum temperature $T\sim10\,$K for the cores, 
disk gas surface density $\Sigma_{\rm gas}\sim10\,\msun\,{\rm pc^{-2}}$, 
and large-scale turbulent velocity dispersion $\sigma_{\rm g}(h)\sim 10\,{\rm km\,s^{-1}}$ 
($\mathcal{M}_{h}\sim30$). For all cases, we will assume a spectrum $p=2$, 
which is what is expected in highly super-sonic simulations and 
suggested by observations of the ISM \citep{burgers1973turbulence,
larson:gmc.scalings}. As discussed above 
and in \papertwo, our results are not qualitatively changed for $p=5/3$ instead. Together this is sufficient 
to completely specify the model.

Next, consider canonical parameters in ULIRG/starburst regions. 
Observations typically find $T\sim60-80\,$K in the dense molecular 
gas ($n\sim 10^{5}-10^{7}\,{\rm cm^{-3}}$) with 
turbulent velocity dispersions $\sigma_{\rm g}(h)\sim40-100\,{\rm km\,s^{-1}}$ 
($\mathcal{M}_{h}\sim50-100$) \citep{downes.solomon:ulirgs,
bryant.scoville:ulirgs.co,westmoquette:m82.sb.core.gasdynamics,
greve:2009.sb.molgas.props}, 
and surface densities $\Sigma_{\rm gas}\sim 10^{3}-10^{4}\,\msun\,{\rm pc^{-2}}$ 
\citep[e.g.][]{kennicutt98}. 

In Fig.~\ref{fig:imf.env}, we compare the CMF predicted in each case.
Although the broad qualitative behavior is similar, there are striking differences. 
As expected from our analysis in \S~\ref{sec:scalings}, the characteristic mass 
in the ULIRG case is somewhat smaller. From Eq.~\ref{eqn:msonic}, 
compare $M_{\rm sonic}\sim 1.6\,\msun$ (MW) to $M_{\rm sonic}\sim 0.4\,\msun$ (ULIRG).
But even if we ignore the shift in the scale of $M_{\rm sonic}$, the low-mass 
turnover occurs much more slowly in the ULIRG case, as expected from 
Eq.~\ref{eqn:slope.lowmass} from the higher 
$\mathcal{M}_{h}$. The high-mass {slope} is nearly identical in both cases, 
as expected from Eq.~\ref{eqn:slope.highmass}, but there is an intermediate-mass steepening 
in the ULIRG case that owes to second-order effects from the run in $S(R)$.

An even more extreme example is the pc-scale circum-BH disk in the MW. 
Here, the initial gas conditions are uncertain; but from the observations, 
we can take the effective radius $R\approx 0.5\,$pc and infer a surface density $\Sigma_{\rm gas}\sim 0.6\epsilon^{-1}\times10^{4}\,\msun\,{\rm pc^{-2}}$ (where $\epsilon = M_{\ast}/M_{\rm gas}$ is 
the unknown star formation efficiency of the progenitor disk).
Simulations of these circum-BH disks in \citet{nayakshin:sfr.in.clumps.vs.coolingrate,
hobbs:2009.mw.nucdisk.sim} suggest an order-unity $\epsilon\sim0.5$ and 
$\sigma_{\rm g}\sim2\,{\rm km\,s^{-1}}$, which is approximately what is needed to 
give $Q\sim1$  (for $M_{\rm BH}=4\times10^{6}\,\msun$, and $\kappa\approx\Omega$ 
for the quasi-Keplerian orbit here). 
We adopt the same typical nuclear gas temperatures as the ULIRG case. 
The CMF in this case is very different from the previous cases. This is because the system 
is extremely thin and $\kappa$ large, dominated by the external BH potential, so $\mathcal{M}_{h}$ is relatively small, but also our assumptions that $R_{\rm sonic}\ll h$ and ${\rm d}B/{\rm d}S\gg 1$ break down and the the correction terms in $\flast$ not captured in Eq.~\ref{eqn:approx} dominate.

\begin{figure}
    \centering
    \plotonesize{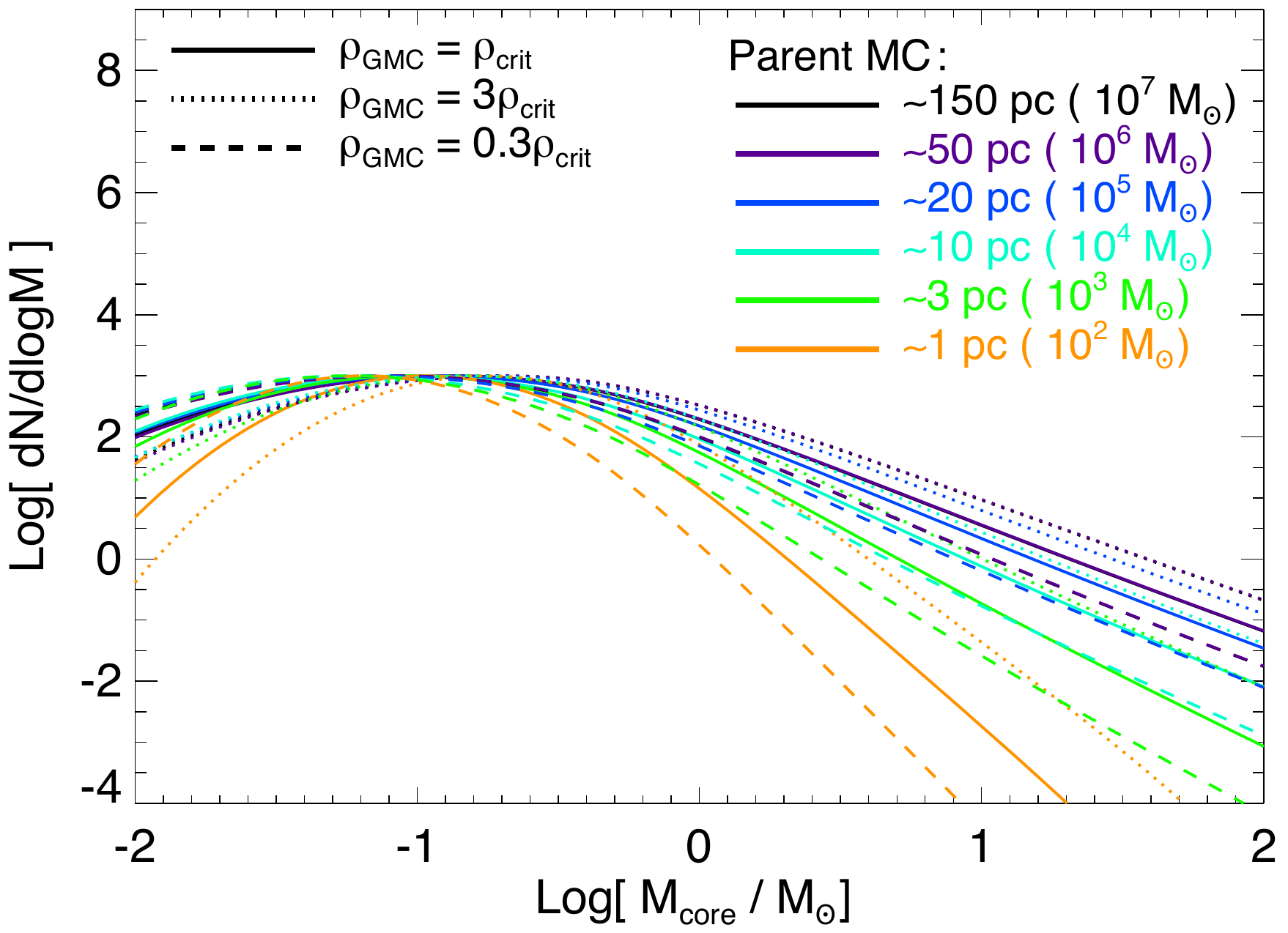}{0.95}
    \caption{The CMF as Fig.~\ref{fig:imf.env}, but for different ``subregions'' of the 
    MW-like system. We compare the CMF in different isolated parent MC masses 
    (the parent largest mass/size 
    scale on which the gas is self-gravitating, i.e.\ has $\langle \rho(R_{\rm cloud})\rangle = \rho_{\rm crit}$), 
    for our standard parameters (solid lines; 
    size and mass of the MC labeled). For convenience, we normalize each model to the same 
    maximum number density. 
    We compare the IMF predicted inside a parent cloud on the same spatial scale but with 
    higher (dotted) or lower (dashed) mean density. Even though clouds have different 
    {\em cloud-scale} Mach numbers ($\langle \mathcal{M}^{2}(R_{\rm cloud})\rangle^{1/2}$ increases 
    as $R_{\rm cloud}^{1/2}$ from $\sim 0.1\,\mathcal{M}_{h}$ to $\sim 1.5\,\mathcal{M}_{h}$ 
    in the smallest to largest clouds here) and density 
    ($\rho_{\rm crit}/\rho_{0}$ decreasing from $\sim50$ to $\sim2.5$),  
    the CMF changes weakly with parent mass, provided it is large enough to properly sample 
    the mass range. The high-mass end is suppressed in the lowest-mass clouds, 
    but mostly because they lack sufficient mass to form many high-mass cores.
    \label{fig:imf.var.subregions}}
\end{figure}

\vspace{-0.5cm}
\section{CMF \&\ IMF Variation in Different Local Regions}
\label{sec:var.local}

The above examples are intentionally extreme. Normal galaxies exhibit a much more narrow range of 
$\Sigma_{\rm gas}$, $T$, and $\sigma_{\rm g}$. 
In Fig.~\ref{fig:imf.var.normal} we compare the CMF predicted for a ``typical'' range in these parameters. 
In all these cases, the high-mass slope is nearly identical and close to the Salpeter value. 
Modest changes in $\mathcal{M}_{h}$ still lead to differences at the low-mass end, although the shape is quite similar. We compare these variations to the observed range inferred for the IMF at these masses in different systems (or equivalently, at these masses, the PDMF); the range in ``turnover speed'' is consistent.\footnote{\label{foot:bastian.imf.corr}Specifically, we compile the MFs from different regions from Fig.~3 in \citet{bastian:2010.imf.universality} and Fig.~1 in \citet{de-marchi:2010.pdmf.cluster.var}, correct them to core masses with the same conversion assumed throughout, normalize them by assuming the total integrated mass over each MF is the same, and then plot as the shaded range the contours that enclose $14-86\%$ ($\pm1\,\sigma$) of the points in each $\approx 0.2\,$dex bin of mass.}

There is also the question of how the CMF/IMF should vary within different MCs or sub-regions of the Galaxy with fixed {\em global} properties. Recall, what we calculated in Figs.~\ref{fig:imf.env}-\ref{fig:imf.var.normal} was the CMF integrated over the entire galaxy. One advantage of our approach is that it {\em automatically} includes the fact that star formation is clustered (see \citealt{hopkins:excursion.clustering}), with correlated star formation inside of identifiable large-scale bound gas clouds (MCs) with a broad spectrum of properties in good agreement with observations (see \paperone). Indeed, by definition, every last-crossing is embedded inside a larger-scale first-crossing, so in calculating the CMF we have implicitly summed over all ``clusters'' (defined loosely here as all regions with cores inside of a parent self-gravitating cloud), the same as is done observationally when summing the contributions from different clusters or integrating galaxy-wide \citep[see e.g.][]{weidner:2006.max.mass.cluster,kroupa:2011.imf.review}. But it is also interesting to ask how the CMF might vary within those specific regions. 

To address this we use the solution to the two-barrier problem, which lets us calculate the CMF for any given overdensity on a larger ``parent'' scale. 
In \paperone, we identified self-gravitating regions on larger scales (which contain sub-regions of fragmentation), i.e.\ crossing the barrier $B(S_{0})$ at smaller $S_{0}<S(M)$, as MCs. 
We can therefore calculate the CMF for a MC of size/mass scale $R_{0}(S_{0})$ or $M_{0}(S_{0})$ 
by taking the two-barrier solution with $\delta_{0}(S_{0})=B(S_{0})$. 
We show this in Fig.~\ref{fig:imf.var.subregions} for a range of ``parent'' MC scales. 
We can further arbitrarily vary $\delta_{0}$ on each scale: we consider regions more/less dense than the ``collapse threshold'' (typical MC) at the same spatial scale. The results are typically less sensitive to parent cloud properties than the other parameters we have considered. Because the {\em global} parameters are fixed, parent scale does not change $\mathcal{M}_{h}$ or the sonic length -- it is largely equivalent to shifting the ``background'' mass normalization. The high-mass end is suppressed in lower-mass MCs, when the MC masses are sufficiently low that they cannot contain the highest-mass cores.

\vspace{-0.5cm}
\section{Discussion}
\label{sec:discussion}

We have used the method from \paperone\ and \papertwo, which developed a 
formal means to calculate the statistical properties of bound objects formed by 
turbulent density fluctuations in a galactic disk, to study the predicted variation in 
the pre-stellar core MF, and by implication the stellar IMF, across different environments. Specifically, in 
\paperone, we showed how the full excursion set formalism can be applied to 
lognormal density distributions in a galactic disk, and demonstrated that, for a 
given turbulent power spectrum shape $E(k)\propto k^{-p}$, their dimensionless properties 
were only a function of the normalization of that power spectrum on large scales 
$\mathcal{M}(h)$. In \papertwo, we extended this to define the ``last crossing distribution'' 
which defines the mass spectrum of self-gravitating cores on the smallest scales on 
which they are bound; following \citet{hennebelle:2008.imf.presschechter} 
we showed that for typical conditions, this 
predicts a mass function in good agreement with observed CMFs and (for plausible core-to-stellar mass conversion efficiencies) consistent with the shape of the canonical stellar IMF. 

Here, we show that under ``normal'' conditions -- specifically, a  
range of star-forming core temperatures $T\sim5-20\,$K, 
galaxy velocity dispersions $\sigma_{\rm g}\sim5-15\,{\rm km\,s^{-1}}$, 
and galaxy surface densities $\Sigma_{\rm gas}\sim 1-20\,\msun\,{\rm pc^{-2}}$ -- 
there is almost no variation in the high-mass (Salpeter) behavior. 
In \paperone\ we showed this is also true if we vary the turbulent power 
spectrum shape $p=5/3-2$. 
The low-mass end is somewhat sensitive to the global Mach number 
$\mathcal{M}_{h}$, turning over more slowly in systems with higher $\mathcal{M}_{h}$ 
(per Eq.~\ref{eqn:slope.lowmass}).
The same result was obtained with a more approximate derivation of the 
CMF in \citet{hennebelle:2008.imf.presschechter,hennebelle:2009.imf.variation}. The change is not especially large, 
however, corresponding to a range of slopes $\alpha\sim 0-1$ at $\sim 0.1\,\msun$ 
(with larger, but much more uncertain, variation below this). 
We stress that this is the {\em global} $\mathcal{M}_{h}$, it does not necessarily 
imply variation in regions with different {\em local} $\mathcal{M}_{h}$ (since their 
density field has contributions from all scales).

To address that question, 
in \citet{hopkins:excursion.clustering}, we extend the excursion set model to derive the solution to the
two-barrier last-crossing 
problem. This corresponds to the solution for the 
CMF in sub-regions of any specified size/mass scale, within a common parent system 
of fixed global properties. We apply this here to examine how the CMF should vary 
in different MC sub-regions within the MW (recall, our galaxy-wide prediction implicitly integrates over the CMF in all individual MCs with the full spectrum of properties over the entire disk). Since in \paperone\ we show that MCs can be identified as solutions to the first-crossing problem, we can 
specifically show how the CMF should vary in ``typical'' MCs of different masses, 
and higher/lower density regions. The variation is minimal in all regions of modest mass 
(despite systematically different densities and local $\mathcal{M}$). 

At low MC masses the CMF is truncated, but this is basically just because there is not enough collapsing mass to form the most massive cores. Since the mass density in MCs is primarily in high-mass systems and this occurs only in the lowest-mass clouds, it has small global effects \citep[but see][]{weidner:2006.max.mass.cluster,weidner:2010.mmax.vs.mcluster}. The mass scale of the predicted CMF does {\em not} vary systematically with parent MC mass.

These predictions agree with the fact that there is remarkably little variation in the high-mass IMF behavior observed in different MW and local group star forming regions. And the predicted low-mass variation in these regions in fact quite similar to the scatter inferred between different regions observationally (although some of this may owe to observational uncertainties). 

If, however, we consider much more extreme systems, we predict significant variations in the CMF. 
For the typical conditions in ULIRGs and other dense, starburst regions, we predict a {\em bottom-heavy} CMF. There is strong observational and theoretical evidence that the stars in the central $\sim$kpc of essentially all ellipticals must have formed in nuclear (probably merger-induced) starbursts, and the properties of recent ``ULIRG/starburst relics'' agree well with elliptical/bulge centers 
\citep[see e.g.][]{kormendysanders92,hibbard.yun:excess.light,rj:profiles,
hopkins:cusps.mergers,hopkins:cusps.ell}. In fact \citet{hopkins:sb.ir.lfs} showed that the 
central densities and mass profile structure of ellipticals imply, if they obeyed any sort of Kennicutt-Schmidt relation in their formation, that they must have been starbursts with densities and velocity dispersions quite similar to what we assume in Fig.~\ref{fig:imf.env}. This may, therefore, explain recent observational suggestions that the central regions of nearby Virgo and other massive ellipticals exhibit a bottom-heavy IMF similar to that predicted here 
\citep{van-dokkum:2010.bottom.imf.ells,van-dokkum:2011.imf.test.gcs,treu:2010.early.type.imf.lensing,cappellari:2012.imf.variation,ferreras:2012.imf.vs.vel.dispersion.early.type.gals,spiniello:2012.bottom.heavy.imf.massive.gal}.

The key reason for this is that these regions have much larger global Mach numbers $\mathcal{M}_{h}$ than the typical MW regions, even if we allow for much larger minimum gas temperatures $T\sim50-100\,$K.
Mathematically, the predicted bottom-heavy CMF in ULIRGs corresponds to the slow low-mass turnover predicted in Eq.~\ref{eqn:slope.lowmass}, although there are higher-order corrections that only appear 
in the exact solution (not the approximate CMF derivation in \citet{hennebelle:2008.imf.presschechter}). 
Physically, this is a result of turbulent fragmentation. Although the characteristic turbulent Jeans mass -- which determines the {\em largest} structures -- increases (scaling $\propto \sigma^{4}$), this appears in the {\em first}-crossing distribution: i.e.\ it means larger MCs/GMCs (in size and mass), not larger cores. 
The MC/first-crossing structures are not directly relevant, since they will be self-gravitating on many smaller scales -- i.e.\ fragment -- internally. But the CMF turnover is not exactly set by the local thermal Jeans mass either (recall, we also assume significantly higher minimum temperatures in this regime). Strictly, the mass scale of the last-crossing distribution is set by the sonic mass, so there is some competition between higher $\mathcal{M}_{h}$ (which drives the sonic length down) and higher $c_{s}$. If the system is globally stable, we can equate this mass scale to purely global parameters, and infer that the rise in surface density $\Sigma_{\rm gas}$ is sufficiently rapid that it ``wins out'' and lowers the sonic mass. 
However, even if this mass scale were the same or somewhat higher, the CMF would still be quite bottom-heavy: a higher $\mathcal{M}_{h}$ not only shifts the sonic mass scale, but also imprints larger density fluctuations in the turbulent field on all scales. This promotes more fragmentation over a more broad range of masses, ``slowing down'' the low-mass turnover or even steepening the slope towards low masses \citep[for simulations demonstrating this, see e.g.][]{ballesteros-paredes:2006.imf.turb.sims}. The same physics may explain suggestions of more bottom-heavy IMFs at higher metallicities \citep[both galactic and extragalactic; see][and references above]{kroupa:1994.bottom.heavy.imf.cooling.flows,kroupa:2001.imf.var,marks:2012.lowmetal.topheavy.imf}, as more efficient cooling can give rise to systematically higher $\mathcal{M}_{h}$.

This is an example where the distinction between the first and last-crossing distributions is critical: if we ignored it, we would arrive at similar conclusions to \citet{padoan:1997.density.pdf} -- essentially, the exact opposite of what we find -- where the authors argued that starburst systems should have more top-heavy IMFs. As well, it is clearly critical to account for the different global turbulent conditions, presumably driven on the largest scales in a galactic disk (and therefore not necessarily captured in many idealized simulations of small volumes), since it is the competition between turbulent density fluctuations and thermal support (not just the latter) that determines the shape of the CMF.

Another extreme system is the $\sim0.5\,$pc-scale circum-BH disk in the MW, which observations have suggested may have a top-heavy IMF \citep{paumard:2006.mw.nucdisk.imf,maness:2007.mw.nucdisk.imf,bartko:2010.mw.nucdisk.imf}. We attempt to extend our estimate to this system as well, although we caution that the characteristic parameters during its formation are quite uncertain and our model may not be applicable at all because of non-linear tidal and eccentricity corrections \citep{alexander:2008.mw.nucdisk.sims}. The predicted CMF is somewhat top-heavy, but it also has an unusual shape which is much more restricted in mass range (truncated by shear and tidal forces). Part of this top-heavy character owes to its having a relatively small $\mathcal{M}_{h}$ if it has $Q\sim1$ (for a $Q=1$ disk near a dominant BH, $\sigma_{\rm g}/V_{c}\sim M_{\rm disk}/M_{\rm BH} \sim 10^{-2}$ here). But it is complicated because the disk has such a small scale height that the distinction between first and last crossings is blurred. Nevertheless, it is suggestive and agrees with results from numerical hydrodynamic simulations \citep{nayakshin:sfr.in.clumps.vs.coolingrate,hobbs:2009.mw.nucdisk.sim}. 

We caution that, as discussed in \citet{hennebelle:2009.imf.variation}, the low-mass end is also the transsonic regime (below the sonic length), so our simple assumption of isothermal gas will break down at some level. Complicated equations-of-state, magnetic fields, radiative feedback, and cosmic-ray heating are certainly capable of modifying the relation between CMF and IMF; in simulations, these add to the thermal support in starburst regions and have motivated arguments for a top-heavy IMF \citep[see e.g.][]{klessen:2000.cluster.formation,klessen:2007.imf.from.turbulence,clarke:2000.starcluster.formation.orion,bate:2005.imf.mass.jeansmass,kroupa:2003.lowmass.frag,jappsen:2005.imf.scale.thermalphysics,papadopoulos:2011.ecrdrs}. Within the context of the present, simplified model, we can only approximately account for the effect of these physics, manifest in their changing the isothermal ``floor'' temperature (which we freely vary). Preliminary comparison of models with more complicated equations of state in \citet{hopkins:frag.theory} suggests this may capture some of the lowest-order effects, if the effective temperature is well-chosen, but this probably remains the largest uncertainty in our model. On the other hand, simulations can only explore a limited parameter space, and owing to severe resolution and timestep demands very few have considered Mach numbers and densities as extreme as observed in real starbursts. Further development of the models, to allow non-linear changes in the thermal state of the gas especially through the time-dependent fragmentation process of contracting cores, would therefore be of great interest. This is critical to address fundamental questions, such as whether (especially at high masses) single cores can be associated with single stars, and (if not) whether we can statistically map the CMF to the IMF at all. For now, our predictions should not be taken too literally as an absolute prediction for the IMF. Rather, our intention is to motivate some potential observed IMF variations (or lack thereof) as a consequence of the variations in the pre-stellar core mass function with global galaxy properties.

\vspace{-0.7cm}
\acknowledgments 
We thank Robert Feldmann, Charlie Conroy, and Mark Krumholz for many helpful discussions during the development of this work, as well as Nate Bastian, Patrick Hennebelle, and the anonymous referee.  Support for PFH was provided by NASA through Einstein Postdoctoral Fellowship Award Number PF1-120083 issued by the Chandra X-ray Observatory Center, which is operated by the Smithsonian Astrophysical Observatory for and on behalf of the NASA under contract NAS8-03060.\\

\bibliography{/Users/phopkins/Documents/work/papers/ms}

\end{document}